\documentclass[apj,iop]{emulateapj}

\usepackage{graphicx}
\usepackage{algorithm}
\usepackage{algorithmic}
\usepackage{amsmath}

\usepackage{color}

\newcommand{\colwidth}{0.475\textwidth}

\slugcomment{In preparation, to be submitted to The Astrophysical Journal}

\begin{document}

\title{A GPU-Computing Approach to Solar Stokes Profile Inversion}
\author{Brian J. Harker}
\affil{National Solar Observatory, Tucson, AZ 85719}
\email{bharker@nso.edu}
\and
\author{Kenneth J. Mighell}
\affil{National Optical Astronomy Observatory, Tucson, AZ 85719}
\email{mighell@noao.edu}

\begin{abstract}
We present a new computational approach to the inversion of solar photospheric
Stokes polarization profiles, under the Milne-Eddington model, for vector
magnetography.  Our code, named {\sc genesis} ({\sc gene}tic {\sc s}tokes
{\sc i}nversion {\sc s}trategy), employs multi-threaded parallel-processing
techniques to harness the computing power of graphics processing units
({\sc gpu}s), along with algorithms designed to exploit the inherent
parallelism of the Stokes inversion problem.  Using a genetic algorithm
({\sc ga}) engineered specifically for use with a {\sc gpu}, we produce
full-disc maps of the photospheric vector magnetic field from polarized
spectral line observations recorded by the Synoptic Optical Long-term
Investigations of the Sun ({\sc solis}) Vector Spectromagnetograph ({\sc vsm})
instrument.  We show the advantages of pairing a population-parallel genetic
algorithm with data-parallel {\sc gpu}-computing techniques, and present an
overview of the Stokes inversion problem, including a description of our
adaptation to the {\sc gpu}-computing paradigm.  Full-disc vector magnetograms
derived by this method are shown, using {\sc solis/vsm} data observed on 2008
March 28 at 15:45 {\sc ut}.\footnote{Full-resolution versions of the images
in this paper are available in the journal- and electronic-version.}
\end{abstract}


\section{INTRODUCTION}\label{sec:intro}
Since \citet{hale:1908} first inferred the presence of magnetic fields on the
sun by observing the Zeeman-induced separation of the components of
magnetically-sensitive spectral lines, the reliable determination of vector
fields from spectral line observations has played a pivotal role in diagnosing
solar magnetism.  The Stokes vector,
${\bf I}_{\lambda}=[I_{\lambda},Q_{\lambda},U_{\lambda},V_{\lambda}]^{T}$,
whose elements are linear combinations of measured polarization intensities at
a wavelength $\lambda$, provide a convenient way to observe and characterize
the effects of magnetic fields on the absorbing medium.  \citet{unno:1956} first
described the Zeeman-induced attenuation of the Stokes vector by magnetic
fields in the framework of the radiative transfer equations neglecting
magneto-optical effects.  The solution was later generalized to include
magneto-optical (Faraday rotation) effects by \citet{rachkovsky:1962}.  

Much information can be directly extracted from the observed Stokes vectors
themselves; \citet{rees:1979} developed a center-of-gravity technique for
estimating the longitudinal component of the magnetic field from Stokes $I$ \&
$V$ observations, and \citet{ronan:1987} described an integral method for
recovering the longitudinal and transverse fields from integrated Stokes linear
and circular polarization profiles.  Furthermore, several convenient weak-field
calculation ``recipes'' can be found in \citet{landideglinnocenti:1994}.

The unique specification of the magnetic and thermodynamic state of the solar
photosphere solely from observations of the Stokes vector is classified as an 
``inverse'' problem.  These types of problems are often computationally complex
and may be ill-conditioned.  Conversely, the forward-modeling of the Stokes
vector emergent from an assumed model atmosphere is trivial.  This asymmetry
can be exploited to interpret the observed Stokes vector within the framework
of a magnetic model atmosphere, by tuning its configuration to minimize (in a
least-squares sense) the model deviation from the observed Stokes vector.  We
collectively refer to all such solution procedures as ``inversion methods.'' 
\citet{auer:1977} developed a now traditional Stokes inversion method based on
the Levenberg-Marquardt ({\sc l-m}) algorithm
\citep{levenberg:1944,marquardt:1963}, which was subsequently extended and
improved upon by \citet{skumanich:1987}.  This optimization approach performs a
nonlinear least-squares fit of a Milne-Eddington model to the observations,
returning the set of atmospheric parameters that describe the polarized
spectral line.  \citet{ruizcobo:1992} extended the optimization method by
creating an inversion code called {\sc sir} ({\sc s}tokes {\sc i}nversion based
on {\sc r}esponse functions), which performs the inversion of observed profiles
while simultaneously inferring the stratification of the model parameters with
optical depth.  

More recently, artificial intelligence methods and pattern recognition
approaches have been gaining ground in the computational methods of
spectropolarimetric analysis.  \citet{carroll:2001} proposed a technique
based on an artificial neural network ({\sc ann}), whereby a large database
of Stokes profiles (either synthetic or pre-inverted by some other means) is
used to train the {\sc ann} to recognize the functional relationship between
the model parameters and the spectral lines in the training set.  Once
suitably trained, the network can generalize the relationship to other samples
not explicitly included in the training set.  \citet{rees:2000} developed a
technique, based on the Principal Component Analysis ({\sc pca}) formulated by 
\citet{pearson:1901}, to decompose a line profile into their so-called
{\it eigenprofiles}.  The eigenvalues associated with these eigenprofiles
define a point in the model manifold, from which the associated model
parameters can be calculated by interpolation over the training set.  This
method has been subsequently used on real observations by
\citet{socasnavarro:2001} and \citet{eydenberg:2005}.

This paper introduces a new approach to the synthesis and inversion of spectral
lines, based on graphics processing units ({\sc gpu}s), which can rapidly
calculate full-disc vector magnetic fields at the photospheric level.  The
technique is based on the combination of a highly-parallel genetic algorithm
with a computing architecture well suited to exploit the many levels of
parallelism in the Stokes inversion problem.  The remainder of this paper is
organized as follows.  Section \ref{sec:data} presents the observational data
used in this work.  Sections \ref{sec:genetic} and \ref{sec:cuda} outline our
genetic algorithm optimization engine and {\sc gpu}-programming techniques,
respectively.  Details of our implementation of Stokes inversion under the 
assumption of a Milne-Eddington atmosphere are given in Section
\ref{sec:implementation}.  Results from the analysis presented in Section
\ref{sec:results}.  Finally, we offer some outlooks on the future
implementation of our method in the vector field data reduction pipeline for
the Vector Spectromagnetograph ({\sc vsm}), the scanning spectropolarimeter
instrument package in operation as part of the Synoptic Optical Long-term
Investigations of the Sun ({\sc solis}) telescope, located at the National
Solar Observatory atop Kitt Peak, AZ.

\section{DATA \& OBSERVATIONS}\label{sec:data}
This work uses full-disc {\sc solis/vsm} observations of the four Stokes $I$,
$Q$, $U$, and $V$ spectra in a 3.45\AA\ bandpass encompassing Fe {\sc i}
multiplet \#816 near 6302\AA\ (see Table \ref{tab:iontable}).  For this work,
we utilize only the absorption line at $\lambda_{0} = 6302.5017$\AA, but note
that our inversion code calculates the full Zeeman pattern of an input
transition.  Laboratory wavelengths for this multiplet and for two nearby
terrestrial O$_{2}$ absorption features are taken from \citet{pierce:1973}.
\begin{table}[!ht]
\begin{center}
\caption{Line-formation parameters for Fe {\sc i} multiplet \#816.}
\label{tab:iontable}
\begin{tabular}{ccccc}
\hline
\hline
Wavelength & Transition & $g_{\mathrm{eff}}$ & $\chi_{e}$ & $\log(gf)$\\
   (\AA)   &            &                    &    (eV)    &\\
\hline
6301.5091 & $^{5}$P$_{2}$--$^{5}$D$_{2}$ & 1.667 & 3.654 & -0.718\\
6302.5017 & $^{5}$P$_{1}$--$^{5}$D$_{0}$ & 2.500 & 3.686 & -1.235\\
\hline
\end{tabular}
\end{center}
\end{table}

\begin{figure}[!ht]
\centering
\includegraphics[width=\columnwidth,clip=TRUE]{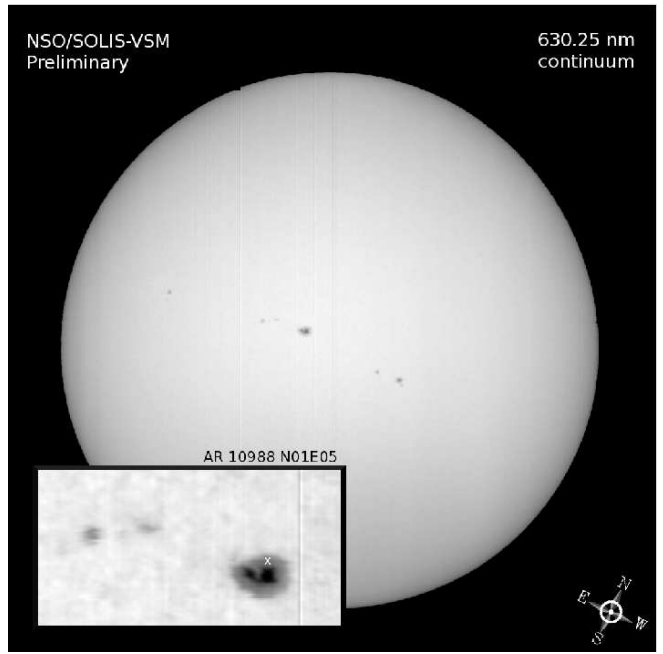}
\caption{Solar disc on 2008 March 28 15:45 {\sc ut} as observed by
         the {\sc solis/vsm} instrument in the 6302.5\AA\ continuum.
         The compass points to heliographic solar north.  The inset
	 shows NOAA 10988 in more detail.}
\label{fig:cont}
\end{figure}

Figure \ref{fig:cont} shows the solar disc as it was observed by{\sc solis/vsm}
in the Stokes $I$ continuum (redward of 6302.5\AA), on 2008 March 28 at 15:45
{\sc ut}.

\begin{figure}[!ht]
\centering
\includegraphics[width=\colwidth,clip=TRUE]{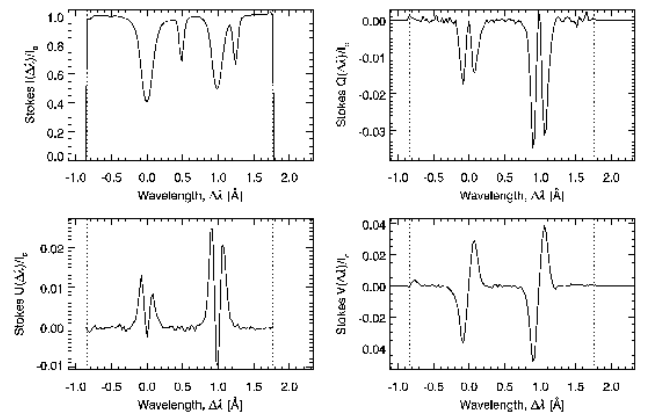}
\caption{Observed Stokes profiles (normalized by the local continuum) for a
	penumbral pixel (represented by the $\times$ symbol in the inset of
	Figure \ref{fig:cont}) belonging to NOAA 10988.  {\sc solis/vsm}
	spectral dispersion is $\Delta\lambda = 27.1$ m\AA, and the wavelength
	scale is relative to the Fe {\sc i} $\lambda$6301.5 line-core
	wavelength.  The narrow, shallow lines are terrestrial O$_{2}$
	absorption features, and are ignored in all analyses.  The vertical
	lines denote the illumination edges of the CCD.}
\label{fig:stokes}
\end{figure}

Figure \ref{fig:stokes} shows sample Stokes profiles taken from a penumbral
region in NOAA 10988 (marked by the $\times$ symbol in Figure \ref{fig:cont}),
close to disc-center.  The somewhat low spatial resolution of {\sc solis/vsm}
pixels (1$\arcsec$.125 pixel$^{-1}$) smears the individual Zeeman components
of the Fe {\sc i} 6302.5\AA\ line, preventing observation of fully resolved
Zeeman lobes, even in sunspot umbrae where the splitting is largest.

For the inversion of this dataset, our routines analyze only those pixels whose
fractional polarization degree,
\begin{equation}
p = \frac{\sqrt{Q_{\mathrm{max}}^{2} + U_{\mathrm{max}}^{2} +
	V_{\mathrm{max}}^{2}}}{I_{c}},
\end{equation}
exceeds a threshold set by the polarimetric noise, defined by
\begin{equation}\label{eqn:sigpol}
\sigma_{P} = P_{\mathrm{thresh}}\frac{\sqrt{\sigma_{Q}^{2} + \sigma_{U}^{2} +
	\sigma_{V}^{2}}}{I_{c}}
\end{equation}
where $\sigma_{QUV}^{2}$ are the variances of the Stokes polarization profiles
in a spectrally-quiet region away from the observed spectral lines, $I_{c}$ is 
the continuum intensity (determined in this same spectral window), and
$P_{\mathrm{thresh}}$ is a proportionality factor.  A value of
$P_{\mathrm{thresh}}$ equal to unity will signal the inversion code to process
all pixels with polarization degree strictly larger than the noise, while a
value of $\sim 10$ typically provides good discrimination between active regions
and surrounding quiet-sun, as observed by {\sc solis/vsm}.  While the spectral
uncertainties will vary somewhat between scanlines and pixels, they are
typically of the order of the {\sc vsm} polarimetric accuracy
($\sim$ 0.1\% of the continuum intensity).  Pixels with polarization degree
less than that given by equation \ref{eqn:sigpol} are inverted under the
weak-field approximation; the longitudinal field strength is calculated from
the separation between the $I{\pm}V$ line-cores, while the field orientation is
calculated from Stokes $Q$, $U$, and $V$ maxima, as in \citet{auer:1977}.  To
perform the inversion, we utilize a class of global optimization algorithms
known as {\it genetic} algorithms.

\section{A STANDARD GENETIC ALGORITHM}\label{sec:genetic}
Genetic algorithms ({\sc ga}s) are a broad class of search and optimization
algorithms which exploit computational analogues to the principles of
evolutionary biology, first proposed by \citet{darwin:1859}.  {\sc ga}s operate
on (potential) solutions which are {\it encoded} into a binary representation.
This requires distinction between the {\it phenotypic} expression (the
parameters of the problem themselves) of the {\it genotypic} representation
(the {\it encoded} parameters).

Genetic algorithms are receiving ever-increasing attention as powerful tools for
search, optimization, and pattern recognition in fields as far-ranging as
engineering design \citep{karr:1999}, job-shop scheduling \citep{davis:1985},
and stock market analysis \citep{mahfoud:1996}; \citet{diver:1997},
\citet{mcintosh:1998}, and \citet{federicoramirez:2002} have applied them to
spectral analysis of nighttime astronomical observations;
\citet{charbonneau:1995} created the {\sc pikaia} genetic algorithm and
demonstrated its utility on several distinct problems of astrophysical
importance, while \citet{lagg:2004} more recently applied the {\sc pikaia}
algorithm to the problem of diagnosing magnetic fields in the upper chromosphere
with the He {\sc i} triplet at 10830\AA.  They are quite robust, and often
succeed where more traditional methods fail (e.g.  multi-modal optimization).

The {\sc ga} solution procedure relies on {\it genetic operators} which are
repeatedly applied to a ``parent'' population of candidate solutions to produce
a new ``offspring'' population of solutions, containing better solutions than
those found in the previous population.  The quantity which determines whether
one solution is better than another is called the {\it fitness}, evaluated for
each candidate solution via a {\it fitness function}.  Here, the fitness
function is a $\chi^2$-like merit score describing model deviations from
observations.  After some number of iterations of this process, the population
will have converged to a small neighborhood around candidate solutions which
have the best fitness scores.  Although a comprehensive review of genetic
algorithms and operators is well beyond the scope of this paper, we present
here a simple overview of  basic {\sc ga} properties.  An $N$-bit binary string
is used to encode the $j^{th}$ model parameter,
\begin{equation}
{\bf b}^{(j)} \equiv \left[b^{(j)}_{0}b^{(j)}_{1}b^{(j)}_{2} \cdots
	 b^{(j)}_{N-1}\right], 
\end{equation}
where $b^{(j)}_{i} = \left[0,1\right]$, for each free parameter of the model.
The convention adopted here is that bit $0$ is the most-significant bit, while
bit $N-1$ is the least-significant bit.  If the $j^{th}$ real-valued free
parameter, ${\bf p}_{j}$, is constrained such that
\begin{equation}
u_{j} \leq {\bf p}_{j} \leq v_{j},
\end{equation}
then each binary string can be decoded into a real floating-point value via
the transformation:
\begin{equation}\label{eqn:decode}
{\bf p}_{j} \equiv \mathcal{E}^{-1}({\bf b}^{(j)}) = u_{j} +
	\frac{v_{j}-u_{j}}{2^{N}-1}\left[\sum_{k=0}^{N-1}{\bf b}^{(j)}_{N-k}
	\times 2^{k}\right].
\end{equation}

The genetic algorithm population therefore consists of $N_{\mathrm{pop}}$ binary
strings of the form:
\begin{equation}\label{eqn:genotype}
\overbrace{ \underbrace{1000010101001001}_{N\ {bits}} \cdots
	\underbrace{1010010100100101}_{N\ {bits}} }^{M{\times}N\ {bits}},
\end{equation}
where $M$ represents the number of free parameters of the model.  The following
list elucidates the notation of Algorithm \ref{alg:ga}, which presents a
pseudocode listing of the basic genetic algorithm functionality.  
\begin{itemize}
\item ${\bf P}_{t}$ denotes the population of candidate solutions at generation
	$t$.  Each (real-valued) candidate solution represents a single
	realization of a plane-parallel, Milne-Eddington {\sc m-e} model of the
	line formation region.  We seek the candidate whose forward-modeled
	Stokes vector shows minimum deviation from the observed profiles.
\item ${\bf G}_{t}$ denotes the population of binary-encoded candidate solutions
	at generation $t$.
\item ${\bf F}_{t}$ denotes the population fitness at generation $t$, and is
	generated by application of the evaluation operator, $\mathcal{F}$, to
	${\bf P}_{t}$.  Better candidate solutions will have smaller fitness
	values.
\item The operator $\mathcal{S}$ represents sampling without replacement.  The
	sampling is stochastically biased such that better candidate solutions
	are (probabilistically) more frequently selected from the population.
\item The operator $\mathcal{E}$ represents the encoding of a candidate solution
	into its binary representation.  An $N$-bit binary string gives a
	numerical resolution of
	$\Delta{p}_{i} = \left(v_{i} - u_{i}\right)/\left(2^{N} - 1\right)$,  
	if $u_{i}$ and $v_{i}$ are the lower and upper bounds, respectively, of
	the $i^{th}$ model parameter.  The corresponding decoding operation
	(equation \ref{eqn:decode}) is denoted by $\mathcal{E}^{-1}$.
\item The operator $\mathcal{R}$ represents pair-wise recombination of the
	population.  Two candidate solutions (selected by $\mathcal{S}$) swap 
	segments of their binary representations to create the representations 
	of two new offspring solutions.
\item The operator $\mathcal{M}$ represents a probabilistic mutation for each
	member of the offspring population.  Each bit on the binary string has a
	small probability of being flipped to its complementary bit.  The 
	mutation operator used in this work was designed to favor local
	exploration by mutating less-significant bits more frequently.
\end{itemize}

\begin{algorithm}[H]
\caption{Pseudocode for a standard genetic algorithm.}
\label{alg:ga}
\begin{algorithmic}[0]
{\tt
\STATE $t = 0$
\STATE Create ${\bf P}_{0}$
\STATE Evaluate ${\bf F}_{0} \gets \mathcal{F}[{\bf P}_{0}]$
\WHILE { ({\bf not} termination-condition) }
    \STATE $t \gets t+1$
    \STATE Select: \ \ \ ${\bf P}_{t} \gets \mathcal{S}[{\bf P}_{t-1},
						        {\bf F}_{t-1}]$
    \STATE Encode: \ \ \ ${\bf G}_{t} \gets \mathcal{E}[{\bf P}_{t}]$
    \STATE Recombine:\ \ ${\bf G}_{t} \gets \mathcal{R}[{\bf G}_{t}]$
    \STATE Mutate: \ \ \ ${\bf G}_{t} \gets \mathcal{M}[{\bf G}_{t}]$
    \STATE Decode: \ \ \ ${\bf P}_{t} \gets \mathcal{E}^{-1}[{\bf G}_{t}]$
    \STATE Evaluate: \ ${\bf F}_{t} \gets \mathcal{F}[{\bf P}_{t}]$
\ENDWHILE
}
\end{algorithmic}
\end{algorithm}

We use {\sc gpu}-computing techniques to offload data-parallel, compute-
intensive calculations with high arithmetic intensity to a massively-parallel
high-speed compute architecture.  For this work, we employ an extension of the
genetic algorithm inversion engine developed and described in
\citet{harker:2009} and a single {\sc nvidia} Tesla C1060 {\sc gpu}.  We
utilize {\sc nvidia}'s Compute Unified Device Architecture ({\sc cuda})
programming interface to handle data transfer and computations the {\sc gpu}.
The next section gives a brief introduction to structured parallel programming
with {\sc cuda}.

\section{THE CUDA PROGRAMMING MODEL}\label{sec:cuda}
The {\sc cuda} programming interface consists of a minimal set of extensions to
the standard C/C++ language which allow a properly-constructed data-parallel
algorithm to execute among the thousands of concurrently running threads
resident on the {\sc gpu}.  This is accomplished by a {\it kernel} function,
which controls the computations and memory accesses to be performed at the
thread granularity.  The kernel function is callable from another user-defined
function (the {\it host} function), which controls data transfers to/from the
{\sc gpu} and is responsible for launching the kernel execution.

A traditional criticism of the application of genetic algorithms to real-world
problems is the fact that they tend to be slow, since they process a potentially
large set of candidate solutions.  The {\sc cuda} programming interface allows
{\sc nvidia gpu}s to execute such data-parallel algorithms, by organizing the
computations into groups of concurrently-executing threads called {\it blocks},
while these blocks are themselves organized into groups of concurrently-
executing {\it grids}.  Mirroring this thread organization in the genetic
algorithm population allows us to write a kernel function that dedicates each
block to the calculations for a specific member of the population, while all
threads in each block are dedicated to the per-wavelength calculations required
by the spectral line synthesis and fitness function evaluation for each
population member.  The power of this approach is evident; in a serial genetic
algorithm, the total number of wavelengths to synthesize (per generational
iteration of Algorithm \ref{alg:ga}) is $N_{\lambda}N_{\mathrm{pop}}$, where
$N_{\lambda}$ is the number of wavelengths spanning the spectral line, and
$N_{\mathrm{pop}}$ is the size of the population.  While this can only be done
one wavelength at a time, one population member at a time for the serial
algorithm, using a {\sc cuda}-capable {\sc gpu} allows all
$N_{\lambda}N_{\mathrm{pop}}$ calculations to be done {\it simultaneously},
constrained only by the physical limitations of the {\sc gpu} hardware itself
(i.e., the maximum possible number of concurrently-executable threads and
total onboard memory).  

While a comprehensive review of the {\sc cuda} architecture is outside the scope
of this paper, we refer the interested reader to the {\sc cuda} Programming
Guide and Software Development Kit ({\sc sdk}), currently available from
{\tt http://www.nvidia.com/getcuda}.

\subsection{Thread \& Memory Hierarchy}\label{sec:thread}
\begin{figure}[!ht]
\centering
\includegraphics[width=\colwidth,clip=TRUE]{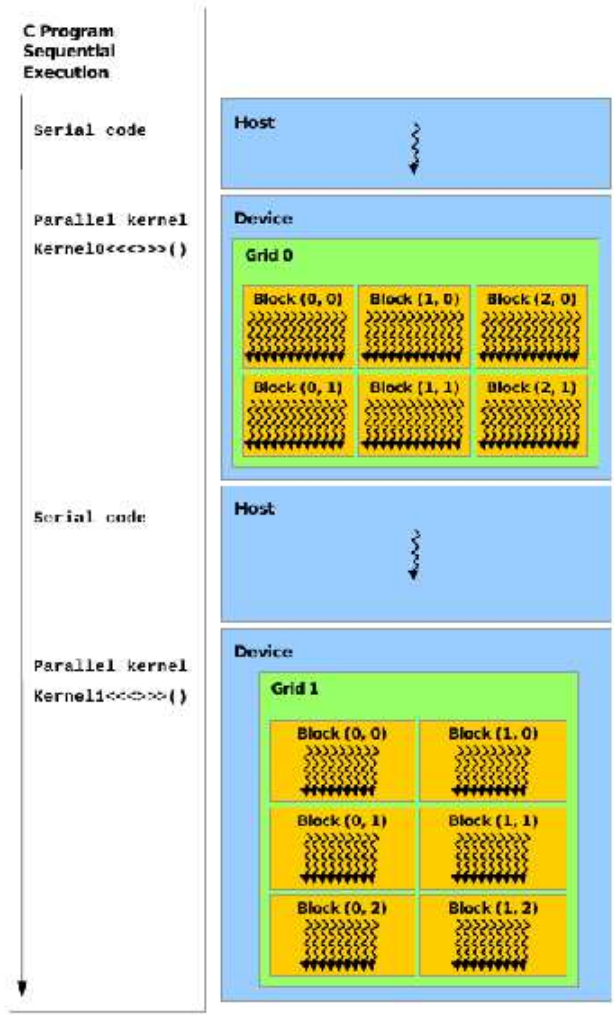}
\caption{{\sc cuda} thread hierarchy ad heterogeneous computing paradigm.
	{\sc cuda} kernel functions are interleaved with the host code, and
	multiple (potentially different) kernels may be launched from a host.
	The $<<<>>>$ syntax is one of the {\sc cuda} extensions to the C
	language; it is used to specify the size and geometry of the execution
	configuration.  Reproduced, with permission, from the {\sc cuda} 
	Programming Guide, courtesy of {\sc nvidia} corporation.
	\citep{nvidia:2011}.}
\label{fig:cuda}
\end{figure}

The {\sc cuda} programming model requires an {\it execution configuration},
whereby the thread distribution is and organized into thread blocks, and
similarly how individual thread blocks are organized into a grid.  Figure
\ref{fig:cuda} shows schematically how an example execution configuration is
indexed, so that each thread can be uniquely identified in the grid.  

Threads from the same thread block may communicate with each other by utilizing
{\it shared} memory (see below), although one must be careful to structure the
program in such a way as to avoid memory access conflicts between threads.
{\it Global} memory allows threads from different blocks to communicate with
each other.  Once the execution configuration is defined, the {\sc cuda} kernel
function is launched by invoking it with a special syntax, shown in Algorithm
\ref{alg:cuda}.  Global memory for the input data and output result is
allocated via {\tt cudaMalloc}, and the data is copied from {\sc cpu} memory
to the allocated {\sc gpu} memory with {\tt cudaMemcpy}.  The kernel function
is invoked with {\tt nBlocks} blocks of {\tt nThreads} threads, and the desired
results are copied back from device to host memory, again via {\tt cudaMemcpy}.
Please note, however, that an actual production-grade kernel invocation is
considerably more involved than the toy example presented in Algorithm
\ref{alg:cuda}.
\begin{algorithm}[H]
\caption{An example C-style {\sc cuda} kernel invocation.}
\label{alg:cuda}
\begin{algorithmic}[0]
{\tt
  \STATE cudaMalloc( (void**)\&dResult, sizeof(hResult) );
  \STATE cudaMalloc( (void**)\&dData, sizeof(hData) );
  \STATE cudaMemcpy( dData, hData, sizeof(hData),
  \STATE \ \ \ \ \ \ \ \ \ \ \ \ cudaMemcpyHostToDevice );
  \STATE MyKernel<<<nThreads, nBlocks>>>( dData, dResult );
  \STATE cudaMemcpy( dResult, hResult, sizeof(hResult),
  \STATE \ \ \ \ \ \ \ \ \ \ \ \ cudaMemcpyDeviceToHost );
}
\end{algorithmic}
\end{algorithm}

\subsection{Memory transfers between {\sc cpu} and {\sc gpu}}\label{sec:memcpy}
An important principle of {\sc gpu}-computing is to minimize the amount of
data that is copied between {\sc cpu} memory and the global memory on the
{\sc gpu}.  Bandwidth across the PCI-Express ({\sc pci-e}) bus connecting
{\sc cpu} memory to {\sc gpu} memory is much lower than the shared memory
bandwidth, as can be seen in Table \ref{tab:bandwidth}.  The table shows the
results of an initial bandwidth test performed on the {\sc gpu} used in this
work; these figures characterize our single realization of hardware components,
and will vary depending on the exact system components and hardware
specifications.  The theoretical maximum bandwidth for the Tesla C1060 is 102
GB s$^{-1}$ \citep{nvidia:2010}, while we achieve $\approx72$\% of this
theoretical limit in our hardware configuration.
\begin{table}[!ht]
\begin{center}
\caption{Initial bandwidth tests for the {\sc nvidia} Tesla C1060 used in
	this work.}
\label{tab:bandwidth}
\begin{tabular}{lc}
\hline
\hline
Transfer Direction              & Bandwidth\\
                                & (GB s$^{-1}$)\\
\hline
{\sc cpu}-to-{\sc gpu} (global) &  1.4634\\
{\sc gpu}-to-{\sc cpu} (global) &  1.1840\\
{\sc gpu}-to-{\sc gpu} (shared) & 73.3165\\
\hline
\end{tabular}
\end{center}
\end{table}

To maximize the efficiency of the data transfer to the {\sc gpu} device, we
modified our genetic algorithm by rephrasing the representation from the
traditional binary arrays to unsigned short integer arrays.  For example,
consider a binary encoding of $M$ parameters with $N=16$ bits each.  Instead
of defining the binary genotype string as \texttt{char gene[M*N]}, with
\texttt{gene[i]} = 0 or 1 (see equation (\ref{eqn:genotype})), the genotype is
defined as \texttt{unsigned short gene[M]}, since each \texttt{unsigned
short} is internally represented by 16 bits.  The former encoding technique
requires $8 \times M \times N$ bits of storage, while the latter requires only
$M \times N$ bits, representing a savings in storage space (and transfer times)
of a factor of 8.  Furthermore, the traditional encoding technique is
incredibly wasteful, since only the least-significant bit of each $8$-bit
\texttt{char} is needed to encode a 0 or 1, meaning only 1/8 of the data
transferred to the {\sc gpu} memory would actually be useful.  In contrast, our
modified binary genotype encodes the same amount of useful information, but
occupies a fraction of the space in memory.  

Transferring the binary encoded parameters to the {\sc gpu} allows the threads
to use the high-speed shared memory to decode the unsigned short integers to
the real floating-point values needed for the model synthesis on the {\sc gpu},
so this is always a net gain in performance over decoding the parameters
serially and transferring the real floating-point values to the {\sc gpu}.

This modified binary genotype also allows our genetic operators to use the
bit-shifting and bit-masking techniques of \citet{iuspa:2001} to operate
directly on the internal binary representation of the unsigned short integers.
These bit-manipulation techniques yield faster genetic operators than would be
used for the traditional binary representation, which typically involve several
nested loops for analysis of each bit in the binary array.  

We have presented the computational aspects of our {\sc gpu} code in this
section, so we now turn to the implementational details of our inversion and
synthesis routines to be run on the {\sc gpu}.

\section{IMPLEMENTATION OF STOKES INVERSION}\label{sec:implementation}
The polarized radiative transfer equations ({\sc prte}) describe the
modification of the Stokes vector of a beam as it propagates, in the direction
$s$, through some medium.  Formally, it is given here as
\begin{equation}\label{eqn:prte}
\mu\frac{d{\bf I}_{\lambda}}{ds} = {\bf K}_{\lambda}\left({\bf I}_{\lambda} -
	 {\bf S}_{\lambda}\right),
\end{equation}
where $\mu$ is the cosine of the heliocentric angle, and ${\bf K}_{\lambda}$
and ${\bf S}_{\lambda}$ are the propagation matrix and source function vector,
respectively.  

If we model both continuum and line absorption processes, then
\begin{equation}
{\bf K}_{\lambda} = {\bf 1} + \eta_{0}{\bf \Phi}_{\lambda},
\end{equation}
where $\eta_{0}$ is the ratio of line-to-continuum absorption coefficients,
and the matrix ${\bf \Phi}_{\lambda}$ includes absorption and magneto-optical
effects, parameterized by the magnetic and thermodynamic properties of the
model atmosphere.  Expressions for these matrix elements may be found in (e.g.)
\citet{landideglinnocenti:2004} and references therein.  These matrix elements
are functions of the Voigt and Faraday-Voigt line profiles, calculated to high
accuracy via the rational function approximation of \citet{hui:1978}.  This
formulation is extremely well-suited to evaluation on a {\sc gpu}, due to its
high arithmetic intensity (ratio of math operations to memory accesses).

The source function vector describes the ratio of emission to absorption in
the beam, and includes both continuum and line contributions, so that
\begin{eqnarray}
{\bf S}_{\lambda} &=& S_{c}{\bf \hat{e}} +
	 \eta_{0}S_{l}{\bf \Phi}_{\lambda}{\bf \hat{e}}, \\
{\bf \hat{e}} &=& \left(1,0,0,0\right)^{T},
\end{eqnarray}
where $S_{c}$ is the continuum source function and $S_{l}$ is the line source
function.  Assuming local thermodynamic equilibrium reduces both continuum and
line source function to the Planck function at the local temperature,
$B_{\lambda}(T)$.  Adopting a Milne-Eddington ({\sc m-e}) relation for the
source function variation as a linear function of optical depth,
\begin{equation}
S_{c} = S_{l} = B_{\lambda}(T) = S_{0} + S_{1}\tau = S_{0}(1 + \beta_{0}\tau),
\end{equation}
where $\beta_{0}=S_{1}/S_{0}$ represents the inverse of the characteristic
length scale over which the source function changes appreciably, the {\sc prte}
admits an analytical solution for the model Stokes profiles,
${\bf I}_{\lambda}^{M}$, given here as
\begin{eqnarray}
\frac{{\bf I}_{\lambda}^{M}}{I_{c}} &=& \left[ \left(1-\beta\right){\bf 1} +
	\beta\left( {\bf 1} + \eta_{0}{\bf \Phi}_{\lambda} \right)^{-1}
	\right]{\bf \hat{e}} \\
\beta &=& \frac{\mu\beta_{0}}{1+\mu\beta_{0}},
\end{eqnarray}
where ${\bf 1}$ is a $4\times4$ identity matrix and $I_{c}$ denotes the observed
local continuum intensity.  This {\sc m-e} solution is characterized
by magnetic and thermodynamic parameters assumed to be constant with
depth through the line-formation region, here collectively represented by the
model vector of free parameters, ${\bf p}$.  Since we do not consider gradients
with respect to optical depth of any parameter except the source function, the
{\sc m-e} atmosphere represents a kind of integrated behavor of the true
parameters over the height of line-formation \citep{orozcosuarez:2007}.
Formally, the $k^{th}$ candidate solution in the genetic algorithm represents
a single model vector 
\begin{equation}
{\bf p}_{k} \equiv [B,\psi,\chi,\lambda_{0},a_{\mathrm{dc}},\Delta\lambda_{D},
	\eta_{0}, S_{0}, \beta_{0}]^{T},
\end{equation}
where $B$ is the magnetic field strength, $\psi$ is the inclination of the
field with respect to the observer's line of sight, $\chi$ is the azimuthal
angle of the field, $\lambda_{0}$ is the line-center wavelength of the
spectral line, $a_{dc}$ is the atomic damping constant of the spectral line,
$\Delta\lambda_{D}$ is the Doppler line-width, $\eta_{0}$ is the
line-to-continuum opacity ratio, and $S_{0}$ and $\beta_{0}$ are the linear
source function coefficients such that the continuum intensity is given by
the Eddington approximation as
\begin{equation}
I_{c} = S_{0} + \mu{S_{1}} = S_{0}\left(1 + \beta_{0}\mu\right),
\end{equation}

To maintain generalizability in the inversion code, the full Zeeman pattern
of the spectral line is calculated at the start of inversion.  This is done
only once, so the overhead incurred is negligible, considering the flexibility
gained.  The particular spectral line to be synthesized is configurable by the
user, and the code contains all the necessary generalizations of the {\sc m-e}
solutions to allow it to function with arbitrary photospheric spectral lines.
The fitness function to be minimized by the genetic algorithm is the following
$\chi^{2}$-like merit function, which quantifies the fit of the Stokes vector
(${\bf I}^{M}$) generated by the model ${\bf p}_{k}$ (with $\nu=4N_{\lambda}-M$
degrees of freedom) to the observations (${\bf I}^{O}$), given here explicitly
as
\begin{equation}\label{eqn:chisq}
\chi^{2}({\bf p}_{k}) = \frac{1}{\nu}\sum_{i} \sum_{j=1}^{N_{\lambda}}
	w^{2}_{ij}\left[{\bf I}^{O}_{i}(\lambda_{j}) -
	{\bf I}^{M}_{i}(\lambda_{j};{\bf p}_{k})\right]^{2},
\end{equation}
where $i=I,Q,U,V$.  The quantities $w_{ij}$ are weighting factors, traditionally
used to adjust the contribution of different wavelengths to the total deviation
across the spectral line, and are discussed further in Section
\ref{sec:weighting}.  Here, the $j$ index is dropped from the weights; they are
taken as constant over the spectral line, but distinct for each of the four
Stokes profiles.  Using this form of the merit function allows the calculation
of uncertainties (over the $\chi^{2}$ hypersurface) in the recovered model
parameters by straightforward techniques, once the genetic algorithm has
converged.

It is an important principle of optimization to work in the smallest possible
parameter space; reducing the dimensionality of the model vector will increase
the speed of the inversion and enhance the stability of the algorithm, since
there are fewer (potentially degenerate) parameters to simultaneously determine.
The next section describes some of the techniques used in our approach to
reduce the dimension of the parameter space and therefore enhance the efficiency
of the genetic search.

\subsection{Reduction of the model manifold}\label{sec:reduction}
Although the line-center wavelength can be a free parameter of the fit,
{\sc genesis} instead directly uses the observed line-center wavelength as
measured from the core of the Stokes $I$ profile.  After calibrating the
observed wavelength scale by measuring the separation of the two terrestrial
O$_{2}$ absorption lines near 6302\AA, a center-of-symmetry approach,
\begin{equation}
\lambda_{\mathrm{sym}} \equiv \underset{\lambda_{i}}{\mathrm{{\bf argmin}}}\
	S(\lambda_{i}) = \displaystyle\sum_{j} \vert I(\lambda_{i+j}) -
	I(\lambda_{i-j}) \vert
\end{equation}
is used to determine a rough estimate of the line-center wavelength.  This
estimate is refined by bracketing $\lambda_{sym}$ with a wavelength triplet
and calculating the minimum of its uniquely-fit polynomial.

The {\sc m-e} model atmosphere specifies a source function linear in optical
depth, characterized by its value at the ${\tau}=0$ photospheric surface
($S_{0}$) and its inverse characteristic length scale ($\beta_{0}$).
Inspection of the Unno-Rachkovsky solutions reveals a simple normalization
scheme that eliminates the dependence on $S_{0}$, as was adopted in
\citet{auer:1977}.  Here, we define a modified Stokes vector,
\begin{equation}\label{eqn:norm}
{\bf I}^{M}_{\lambda} \leftarrow I_{c}{\bf \hat{e}} - {\bf I}_{\lambda}^{M},
\end{equation} 
which consists of the Stokes $I$ {\it line depression} and Stokes $Q$, $U$,
and $V$ profiles.  Note we choose to leave $\beta_{0}$ (which influences the
amplitude of the synthesized Stokes profiles) as a free parameter of the fit.

\subsection{Limited spatial resolution \& scattered light}\label{sec:resolution}
To account for limited spatial resolution, a new free parameter is introduced;
the magnetic fill-fraction $\alpha$ represents the fractional pixel area
occupied by the magnetic field.  The remainder of the pixel area ($1-\alpha$)
is assumed to be field-free.  The Stokes vector then becomes a linear
superposition of the magnetic and non-magnetic profiles, weighted by $\alpha$,
\begin{equation}
{\bf I}^{M}_{\lambda} \leftarrow \alpha{\bf I}^{M}_{\lambda} +
	(1-\alpha)I^{\mathrm{nm}}_{\lambda}{\bf \hat{e}}.
\end{equation}
The non-magnetic profile is assumed to be a quiet-sun Stokes $I$ profile from
the local surroundings.  Using a locally-averaged quiet-sun profile
would require breaking one of the most advantageous properties of the
algorithm, namely that each scanline/pixel can be inverted independently of
the data from neighboring scanlines/pixels.  Furthermore, this approach
requires a tremendous amount of disk I/O, and can be quite slow.  To maintain
a totally independent scanline inversion, we have taken a different approach;
the center-to-limb variation (CLV) of the quiet-sun Stokes $I$ profile
has been measured (Harvey, 2010, private communication) and parameterized as a
pure Voigt function characterized by its amplitude, full-width at half-maximum
(FWHM) and atomic damping parameter.  Gaussian and Lorentzian components of
the profile can be derived from the FWHM.  A 3$^{rd}$-order polynomial is fit
(as a function of heliocentric $\mu$) to the CLV of each of these parameters.
The resulting polynomial fits are used within the inversion to calculate the
appropriate values of the quiet-sun parameters as a function of disc position
for each pixel.  The non-magnetic profile is then synthesized from these
parameters, centered on the observed Stokes $I$ line-center wavelength.

We account for scattered light at the $\sim$5\% level by first correcting the
measured continuum.  The baseline of the observed Stokes $I$ profile is
increased, and subsequently renormalized to the corrected continuum, following
\citet{gray:2005}.

\subsection{Thermodynamic parameters}\label{sec:thermo}
It is well known that there exists some level of degeneracy between the magnetic
and thermodynamic parameters, with respect to their influence on the model
profile line-shapes.  Figure 11.1 of \citet{deltoroiniesta:2003} shows an
explicit example of this effect; it is not always clear whether one combined
magnetic and thermodynamic configuration leads to a better fit to the 
observations than another, even if the configurations are noticeably different
\citep{borrero:2011}.  The convention typically adopted by inversion
practitioners is that the thermodynamic model parameters are of less importance
to the final quality of the fit than the magnetic parameters. 
\citet{skumanich:1987} suggested that in order to find a robust fit, some of the
thermodynamic parameters must be fixed prior to the inversion, and
\citet{borrero:2011} investigated the effects of holding the atomic damping at a
constant value during their inversions.  They found negligible differences
between the recovered vector magnetic fields.  It is not clear, however, that
this procedure is generally acceptable for observations with much higher
spectral resolution than in \citet{borrero:2011}.  We have decided not to hold
fixed any of the thermodynamic parameters, and have implemented a
``pre-fitting'' initialization in which we fit a non-magnetic Stokes $I$ profile
to the observed Stokes $I$ profile, using a simple Levenberg-Marquardt
algorithm.  This returns values for the atomic damping parameter,
$a_{\mathrm{dc}}$, Doppler width $\Delta\lambda_{D}$, line-to-continuum
opacity ratio, $\eta_{0}$, and source function parameter, $\beta_{0}$.
Assuming a non-magnetic model for a (potentially) Zeeman-broadened profile
will, of course, lead to errors in the derived thermodynamic variables.
However, these values are utilized only to constrain the parameter space to
sensible ranges within which the {\sc ga} can search.

\subsection{Weighting scheme}\label{sec:weighting}
Since the Stokes $I$ profile will always have much larger signal strengths than
the polarization profiles, deviations between the observed and synthesized
Stokes $I$ profiles will dominate the $\chi^{2}$ value, essentially causing the
algorithm to fit the model atmosphere solely to the Stokes $I$ intensity
profile.  To mitigate this effect, we equalize the importance of all four Stokes
profiles by using an ``inverse-max'' weighting scheme to ensure that all
deviations contribute roughly equally to the calculated $\chi^{2}$.  The
weighting scheme is given here explicitly as:
\begin{eqnarray}
w_{I} & = & \left( I_{c} - I_{0}^{\mathrm{obs}} \right)^{-1}		    \\
w_{Q} & = & \left( \max\ \vert Q^{\mathrm{obs}}_{\lambda} \vert \right)^{-1} \\
w_{U} & = & \left( \max\ \vert U^{\mathrm{obs}}_{\lambda} \vert \right)^{-1} \\
w_{V} & = & \left( \max\ \vert V^{\mathrm{obs}}_{\lambda} \vert \right)^{-1},
\end{eqnarray}
where $I_{0}^{\mathrm{obs}}$ is the observed line-core intensity.

\subsection{Model manifold boundaries}\label{sec:boundaries}
The genetic inversion will be most efficient in a parameter space that has the
smallest physically-realistic domain for each parameter.  By seeding the
initial population of the genetic algorithm with some specific heuristic
knowledge of the problem at hand, we can both ensure that we start with at least
a few high-quality solutions, and suitably restrict the domain of the searchable
parameter space.  Both scenarios will accelerate the convergence and improve the
final accuracy of the solutions.

Here, the initial population includes representations of magnetic field vectors
generated from the weak-field approximations in \citet{auer:1977}, as well as
from a functional relationship between the field geometry and integrated
measures of the Stokes polarization profiles found in \citet{ronan:1987}.  The
longitudinal field strength estimated from the center-of-gravity approach by
\citet{rees:1979} is also seeded into the initial population.  In the case of
full-disc inversions, where every on-disc pixel is inverted, the trivial
non-magnetic solution is seeded into the population as well.

The field inclination domain is naturally restricted to a single polarity,
$\psi \in [0,90]^{\circ}$ or $[90,180]^{\circ}$, initialized to be in agreement
with the order of the blue- and red-lobes of the observed Stokes $V$ profile.
The seed fields are checked to ensure that they are consistent with this
polarity.

\begin{figure}[!ht]
\centering
\includegraphics[width=\colwidth,clip=TRUE]{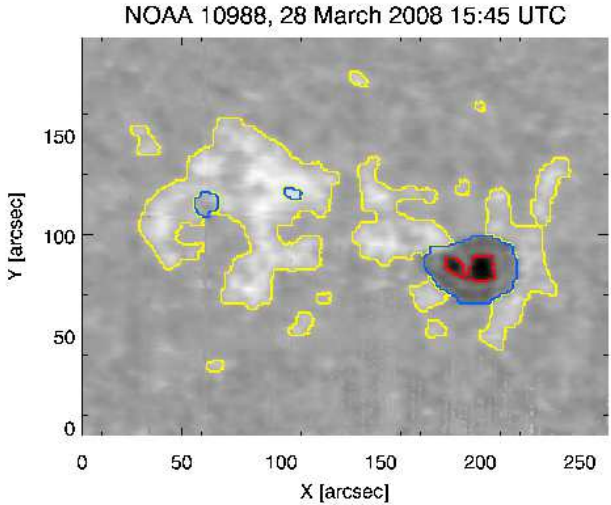}
\caption{Line morphology classification for NOAA 10988.  The image depicts
	Stokes $I$ line-core intensity, overlaid with red, blue, and yellow
	contours demarcating umbral, penumbral, and plage regions,
	respectively.}
\label{fig:regions}
\end{figure}

Additionally, we use empirical knowledge derived specifically from
{\sc solis/vsm} data to discriminate between different magnetic structures
observed on the disc.  Spatial pixels are labeled as likely belonging to the
structures seen in Figure \ref{fig:regions}, based on their observed continuum
and line-core intensities relative to the average quiet-sun profile.  Following 
\citet{hestroffer:1998}, we apply a simple correction for limb-darkening to
flatten the observed intensity profile in pixels far from disc-center, before
a label is assigned.  The details of this empirical classification scheme (and
the subsequent compartmentalization of the $B$-$\alpha$ subspace) are given in
Table \ref{tab:regions}.
\begin{table}[!ht]
\begin{center}
\caption{Active region structure discrimination parameters.}
\label{tab:regions}
\begin{tabular}{cccc}
\hline
\hline
discriminator      & Umbra       & Penumbra    & Plage\\
\hline
$I_{c}/I_{c}^{qs}$ & $<$ 0.65    & [0.65-0.90] & $>$ 0.90\\
$I_{0}/I_{0}^{qs}$ & $<$ 0.69    & $>$ 0.69    & $>$ 0.85\\
$B\ [G]$	   & [1000-3500] & [0-2500]    & [0-1500]\\
$\alpha$           & [0.25-1]    & [0.25-1]    & [0-0.5]\\
\hline
\end{tabular}
\end{center}
\end{table}

\subsection{Population initialization}\label{sec:initialization}
As described in Section \ref{sec:genetic}, the genetic algorithm works by
continually evolving good solutions out of a population of candidate solutions,
represented by the reduced model vectors
\begin{equation}
{\bf p}_{k} \equiv [B,\psi,\chi,a_{\mathrm{dc}},\Delta\lambda_{D},\eta_{0},
	\beta_{0},\alpha]^{T},
\end{equation}
with the number of free parameters, $M=8$.  Hardware limitations dictate the
maximum size of the population; the Tesla C1060 {\sc gpu} contains 30
multiprocessors, each of which is composed of 8 stream processors.  Therefore,
the total number of thread blocks (candidate solutions) which can be
simultaneously processed is $N_{\mathrm{pop}} = 30 \times 8 = 240$.

It is traditional to initialize the genetic algorithm with a random (but
bounded) population to ensure enough diversity for the genetic operators to
produce meaningful evolution.  However, we instead generate the initial
population (of non-seeded candidate solutions) by repeatedly sampling from a
Sobol sequence generated via the \citet{bratley:1988} algorithm.  An
$M$-dimensional Sobol sequence is a {\it quasi}-random sequence that is
maximally self-avoiding; the points in the sequence tend to (roughly) evenly
distribute themselves throughout the $M$-dimensional hypercube $[0,1]^{M}$.
This property gives robust, even coverage of the parameter space without
placing the initial population on a regularized grid, which would completely
inhibit the search action of the recombination operator ($\mathcal{R}$).  In
addition, this approach eliminates chance clustering in the initial population,
thereby avoiding the processing of redundant candidate solutions.  The final
benefit of using quasi-random initialization lies in the efficiency of
population restarts; when the population has converged to some self-monitored
degree, all but the best individual(s) are re-initialized.  Generating new
candidates according to a quasi-random schedule guarantees that we will be
refreshing the ``gene pool'' in the most efficient way, by using the
self-avoidance property to automatically ignore previously-sampled regions
of the parameter space.  We exploit this property to address any issues
related to premature/false convergence; every $N_{\mathrm{samp}}$ iterations
of the genetic algorithm, we discard the worst half of the population (``dead
solutions'') and reinitialize them according to the Sobol mechanism described
above.  In practice, for a maximum number of generations $N_{\mathrm{gen}}$
over which to evolve, we note that $N_{\mathrm{samp}}\approx N_{\mathrm{gen}}/4$
provides a good balance between deep genetic search and the introduction of
new genetic material.  We allow the population to evolve for
$N_{\mathrm{gen}} = 100$ generations.

\section{RESULTS \& DISCUSSION}\label{sec:results}
\begin{figure*}[!ht]
\centering
\includegraphics[width=\textwidth,clip=TRUE]{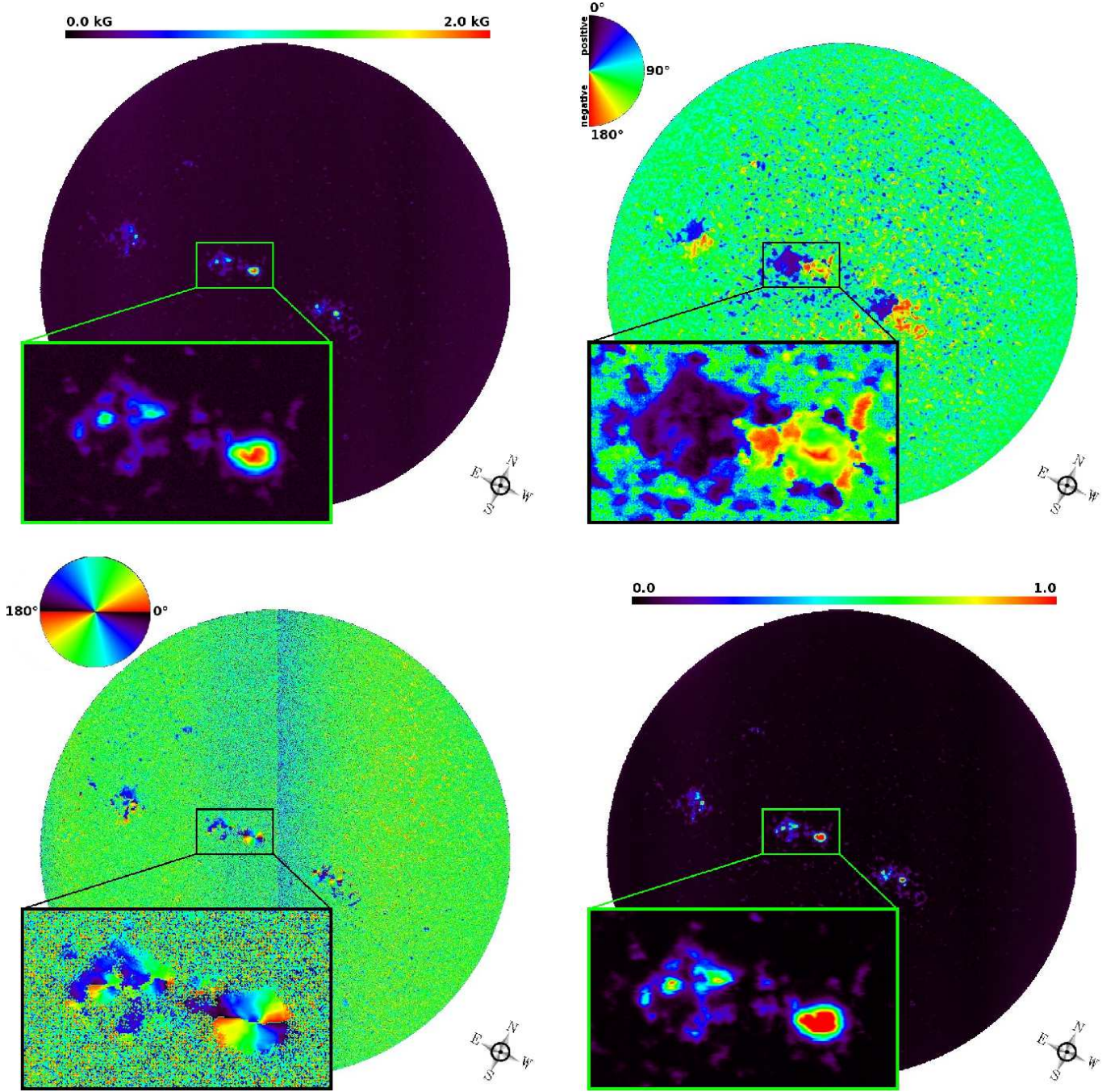}
\caption{{\it (top left)}: field strength, {\it (top right)}: field inclination,
	{\it (bottom left)}: field azimuth, {\it (bottom right)}:
	fill-fraction.  The vertical stripe down the center of the azimuthal
	angle image is an artifact of the dual Rockwell camera system in
	{\sc solis/vsm}, and do not appear in newer Sarnoff camera data.  The
	compass points to solar north.}
\label{fig:pmaps}
\end{figure*}

Proceeding along each scanline and performing the {\sc genesis} inversion on the
corresponding spectra for each pixel builds a map of the model parameters over
the full-disc.  Figure \ref{fig:pmaps} shows the magnetic field strength,
inclination, azimuthal angle, and fill-fraction over the full-disc as inferred
by the {\sc genesis} inversion.  The inset shows NOAA 10988.  

The $0^{\circ}$ reference direction for the azimuthal angle is along the
horizontal axis of the image.  The field has not been resolved of the
$\pi$-ambiguity inherent in all Stokes inversion techniques, hence the
antisymmetric color wheel Nevertheless, the radial structure of penumbral
fields is well-determined.  The fill-fraction displays the expected behavior
of values $\approx$ 1 for the umbral and penumbral regions, with a decline
to values $\leq$ 0.5 for surrounding plage regions.

The statistical spread of the final population provides a convenient means to
generate initial estimates of the uncertainties associated with the fitted
model parameters.  Measuring the spread around the identified optimum gives
{\it population} uncertainties, from which we estimate the gradient of the
$\chi^{2}$ manifold:
\begin{equation}
\left[\nabla\chi^{2}\left({\bf p}^{\mathrm{opt}}\right)\right]_{j} \approx 
	\frac{\chi^{2}\left({\bf p}^{\mathrm{opt}}\right) -
	\chi^{2}\left({\bf p}_{i}\right)}{\left[\Delta{\bf p}\right]_{j}},
\end{equation}
where ${\bf p}_{i}$ is selected from the best of the final population, chosen
such that we avoid numerical difficulties in the calculation of the finite
differences (i.e. ratio of two very small numbers).  Using this estimate, we
bootstrap the derivatives to higher accuracy by Richardson extrapolation
to zero stepsize within Neville's algorithm \citep[see, e.g.,][]{press:1988}.
The Hessian matrix, ${\bf H}$, is thus calculated as the Jacobian of the
resulting gradient vector and the variance of the $i^{th}$ model parameter is
given by \citet{sanchezalmeida:1997}:
\begin{equation}
\sigma_{\mathrm{p}_{i}}^{2} = \left[{\bf H}^{-1}\right]_{ii}
	\frac{\chi^{2}({\bf p}^{\mathrm{opt}})}{M}.
\end{equation}
This evaluation requires a (potentially) large number of fitness function calls,
but is only done once per pixel, so the small overhead incurred is outweighed
by the benefit of having an accurate error estimate for the model parameters.

\begin{table}[!ht]
\begin{center}
\caption{Distribution of uncertainties in Fe {\sc i} $\lambda6302.5$ magnetic
	parameters, broken down by structure.}
\label{tab:errors}
\begin{tabular}{lcccc}
\hline
\hline
Structure & $\Delta_{B}\pm\sigma_{\Delta_{B}}$ &
	    $\Delta_{\gamma}\pm\sigma_{\Delta_{\gamma}}$ &
	    $\Delta_{\chi}\pm\sigma_{\Delta_{\chi}}$ &
	    $\Delta_{\alpha}\pm\sigma_{\Delta_{\alpha}}$\\
          &           (G)              &         ($^{\circ}$)         &
	         ($^{\circ}$)          &     ($\times 10^{-3}$)\\
\hline
Umbra    & 1.3$\pm$2.9 & 0.45$\pm$0.69 & 0.73$\pm$0.89 & 2.4$\pm$5.9\\
Penumbra & 2.4$\pm$4.5 & 0.62$\pm$0.80 & 0.73$\pm$0.91 & 2.5$\pm$5.4\\
Plage    & 1.2$\pm$2.7 & 0.65$\pm$0.79 & 0.84$\pm$0.96 & 2.1$\pm$3.6\\
\hline
\end{tabular}
\end{center}
\end{table}

Table \ref{tab:errors} presents an estimation of the uncertainties in the fitted
model parameters determined in this manner.  The uncertainties are broken
down according to the structure discrimination method in Section
\ref{sec:boundaries}.  For each structure, we calculate the average uncertainty,
$\Delta_{i}$, and corresponding standard deviation, $\sigma_{\Delta_{i}}$.  The
table shows the quantity $\Delta_{i} \pm \sigma_{\Delta_{i}}$, so that
$\Delta_{i} + 3\sigma_{\Delta_{i}}$ represents an upper limit to the model
parameter uncertainties in 99.73\% of the pixels belonging to each active
region structure.  For example, 99.73\% of umbral pixels have uncertainties in
field strength of 10 G or less.  These quantities are not meant to represent
the uncertainties in any particular pixel or structure, but instead
characterize the overall {\it distribution} of uncertainties derived from the
inversion.  It should be noted, however, that these uncertainties are derived
solely from the topology of the $\chi^{2}$ hypersurface, and do not include
systematic errors resulting from the assumptions and/or limitations of the
Milne-Eddington model, which is sometimes not a suitably accurate description
of the real solar atmosphere.

\begin{table*}[!H]
\begin{center}
\caption{Timing profiles for {\sc genesis} inversions of Fe {\sc i}
	$\lambda\lambda$6301.5,6302.5}
\label{tab:timing}
\begin{tabular}{lcccc}
\hline
\hline
mode & \multicolumn{2}{c}{$\Delta T_{\mathrm{run}}$} &
       \multicolumn{2}{c}{$\Delta T_{\mathrm{eval}}$}\\
     & \multicolumn{2}{c}{(min)} &
       \multicolumn{2}{c}{(min)}\\
\hline
               & 6301.5\AA       & 6302.5\AA
	       & 6301.5\AA       & 6302.5\AA\\
\cline{2-5}
{\sc serial}   & 187.83$\pm$0.69 & 84.56$\pm$0.26
	       & 147.54$\pm$0.68 & 55.53$\pm$0.17\\
{\sc gpu-acc.} &  36.40$\pm$0.15 & 32.82$\pm$0.36
	       &   4.43$\pm$0.01 &  2.34$\pm$0.01\\
\hline
\end{tabular}
\end{center}
\end{table*}

Table \ref{tab:timing} shows a speed comparison between the serial and
{\sc gpu}-accelerated versions of the {\sc genesis} inversion code, for both
iron lines, averaged over 50 independent runs.  The total evaluation time
($\Delta T_{\mathrm{eval}}$) is the time spent synthesizing the model spectra
and evaluating the various contributions to equation \ref{eqn:chisq}.  For the
serial code, the evaluation time consumes a strong majority (65.7\% and 78.5\%
for Fe {\sc i} $\lambda$6302.5 and Fe {\sc i} $\lambda$6301.5, respectively) of
the total algorithm runtime, $\Delta T_{\mathrm{run}}$.  Use of the {\sc gpu}
as a co-processor has reduced the total evaluation time to only 7.1\% and
12.1\%, respectively, of the {\sc gpu}-accelerated total runtime.  The total
runtime is quite stable for both the serial and {\sc gpu}-accelerated versions
of the code.  The table shows $1\sigma$ standard deviations, which demonstrate
that the serial and {\sc gpu}- accelerated versions can vary in their total
runtimes by up to a minute.  The {\sc gpu}-accelerated evaluation time further
shows why the {\sc gpu} chip architecture is so amenable to parallel
computations.  The variation in the average time spent interacting with the
{\sc gpu} is less than a second; this stability is precisely due to the
many-cored nature of the {\sc gpu}, which uses a very efficient thread
scheduler to keep the cores of each multiprocessor optimally busy.  

Overall, the {\sc gpu}-accelerated Fe {\sc i} $\lambda$6302.5 inversion is a 
factor of 2.6 times faster than the serial version, with an increase to a factor
of 5.2 for the Fe {\sc i} $\lambda$6301.5 non-normal triplet. The total 
synthesis and evaluation time shows just how powerful the {\sc cuda} programming
paradigm can be, performing the computations 23.7 and 33.3 times faster than
the serial versions of the code can manage.  Since the non-normal triplet Fe
{\sc i} $\lambda$6301.5 has four contributions (from levels with equal
$\Delta{M_{J}}$) to each of the three Zeeman components, a greater proportion
of work is done on the {\sc gpu}.  This highlights another principle of
{\sc gpu} computing; the more work you assign to the {\sc gpu}, the more
efficiently it can perform it.

\section{CONCLUSIONS}\label{sec:conclusions}
We have described a novel computational approach to the inference of
photospheric vector magnetic fields from observations of the Stokes
polarization profiles.  Our new inversion code, named {\sc genesis}, is
capable of quickly producing full-disc spatial maps of the magnetic structure
of the solar photosphere observed by the {\sc solis/vsm} instrument located
atop Kitt Peak at the National Solar Observatory, in a fraction of the time
required by a similar serial technique.  The inversion code is capable of
recovering magnetic fields with uncertainties on the order of 0.5\%, with
errors in the field orientation of a few degrees.  Fill-fractions are
recovered with uncertainties on the order of 2\%.  Currently, the code is only
capable of inverting a single line at a time, though we plan to investigate the
extension to the simultaneous inversion of both lines of the Fe {\sc i}
$\lambda$6302 multiplet.

We have shown the technique to be amenable to the reduction and analysis of
large volumes of spectropolarimetric data.  To this end, we are currently
investigating the the assimilation of the {\sc gpu} hardware and specialized
{\sc cuda}-based algorithms into the {\sc solis/vsm} vector field pipeline.
Our long-term goals for this work are to provide near real-time vector magnetic
fields to the scientific community.  Increasing the cadence of {\sc solis/vsm}
vector data products will also allow us to support and complement observations
taken by the Solar Dynamics Observatory ({\sc sdo}) Helioseismic and Magnetic
Imager ({\sc hmi}), which produces full-disc vector magnetograms at
$4096 \times 4096$ resolution with a cadence of approximately 12 minutes.

The {\sc gpu} programming paradigm is highly scalable; a compiled
{\sc cuda} application can execute on any {\sc cuda}-capable device, subject
to hardware limitations.  The thread scheduler will automatically allocate
the appropriate number of thread blocks to the stream processors.  Coupled with
the Message Passing Interface ({\sc mpi}) to parallelize over scanlines (i.e.
independent scanlines are inverted independently, utilizing their own distinct
{\sc gpu}), this could lead to incredibly fast (i.e. near-realtime or realtime)
full-disc inversions with a modest number of {\sc cpu-gpu} pairs.
Improvements in {\sc gpu} hardware are steadily advancing; the Fermi
architecture is the recent successor to the Tesla architecture, offering up to
512 {\sc cuda} cores, larger-capacity memory banks, and increased
floating-point performance.  With the next-generation Kepler architecture on
the horizon, the prospects for accelerated solar data processing are indeed
promising.  Finally, as {\sc gpu} computing matures, we expect to extend this
approach to better hardware, with a cautious eye toward spectropolarimetric
analyses of data recorded by the upcoming Advanced Technology Solar Telescope
({\sc atst}).  The volume of data expected from this next-generation
observatory will greatly exceed that of the current generation, requiring new
and faster techniques to properly handle and reduce the observations in a
timely manner.  We feel the integration of {\sc gpu}-accelerated data-reduction
techniques will be key for the analysis of such large datasets, and may make
available important (near) real-time information on the photospheric vector
magnetic field to the space-weather forecasting community.

\acknowledgments
The authors wish to thank J. Enos and V. Kindratenko of the National Center for
Supercomputing Applications ({\sc ncsa}) at the University of Illinois at
Urbana-Champagne for kindly providing access to the Accelerator Cluster.  We
are also indebted to H. Lin (Institute for Astronomy) for providing {\sc gpu} 
hardware with which to work locally.  The authors also wish to thank A.
Pevtsov, J. Harvey, and the anonymous referee for helpful comments on the
manuscript.  {\sc solis/vsm} data used here are produced cooperatively by
{\sc nsf}/{\sc nso} and {\sc nasa}/{\sc lws}.  Support for this work was
provided by {\sc nasa} Grant NNH08AH25I (A. Norton, PI).

{\it Facilities:} \facility{SOLIS (VSM)}.

\bibliographystyle{apj}
\bibliography{hark0509}

\begin{thebibliography}{40}
\expandafter\ifx\csname natexlab\endcsname\relax\def\natexlab#1{#1}\fi

\bibitem[{{Auer} {et~al.}(1977){Auer}, {House}, \& {Heasley}}]{auer:1977}
{Auer}, L.~H., {House}, L.~L., \& {Heasley}, J.~N. 1977, \solphys, 55, 47

\bibitem[{{Borrero} {et~al.}(2011){Borrero}, {Tomczyk}, {Kubo},
  {Socas-Navarro}, {Schou}, {Couvidat}, \& {Bogart}}]{borrero:2011}
{Borrero}, J.~M., {Tomczyk}, S., {Kubo}, M., {et~al.} 2011, \solphys, 273, 267

\bibitem[{Bratley \& Fox(1988)}]{bratley:1988}
Bratley, P., \& Fox, B.~L. 1988, ACM Trans. Math. Softw., 14, 88

\bibitem[{{Carroll} \& {Staude}(2001)}]{carroll:2001}
{Carroll}, T.~A., \& {Staude}, J. 2001, \aap, 378, 316

\bibitem[{{Charbonneau}(1995)}]{charbonneau:1995}
{Charbonneau}, P. 1995, \apjs, 101, 309

\bibitem[{{Darwin}(1859)}]{darwin:1859}
{Darwin}, C. 1859, {On the Origin of Species by Means of Natural Selection}
  ({London, U.K., W. Clowes and Sons})

\bibitem[{{del Toro Iniesta}(2003)}]{deltoroiniesta:2003}
{del Toro Iniesta}, J.~C. 2003, {Introduction to Spectropolarimetry}
  ({Cambridge, UK: Cambridge University Press, April 2003.})

\bibitem[{{Diver} \& {Ireland}(1997)}]{diver:1997}
{Diver}, D.~A., \& {Ireland}, D.~G. 1997, Nuclear Instruments and Methods in
  Physics Research A, 399, 414

\bibitem[{{Eydenberg} {et~al.}(2005){Eydenberg}, {Balasubramaniam}, \&
  {L{\'o}pez Ariste}}]{eydenberg:2005}
{Eydenberg}, M.~S., {Balasubramaniam}, K.~S., \& {L{\'o}pez Ariste}, A. 2005,
  \apj, 619, 1167

\bibitem[{{Gray}(2005)}]{gray:2005}
{Gray}, D.~F. 2005, The Observation And Analysis Of Stellar Photospheres
  (Cambridge University Press)

\bibitem[{Grefenstette(1985)}]{davis:1985}
Grefenstette, J.~J., ed. 1985, Proceedings of the 1st International Conference
  on Genetic Algorithms, Pittsburgh, PA, USA, July 1985 (Lawrence Erlbaum
  Associates)

\bibitem[{{Hale}(1908)}]{hale:1908}
{Hale}, G.~E. 1908, \apj, 28, 315

\bibitem[{{Harker}(2009)}]{harker:2009}
{Harker}, B.~J. 2009, PhD thesis, Utah State University

\bibitem[{{Hestroffer} \& {Magnan}(1998)}]{hestroffer:1998}
{Hestroffer}, D., \& {Magnan}, C. 1998, \aap, 333, 338

\bibitem[{Hui {et~al.}(1978)Hui, Armstrong, \& Wray}]{hui:1978}
Hui, A., Armstrong, B., \& Wray, A. 1978, Journal of Quantitative Spectroscopy
  and Radiative Transfer, 19, 509

\bibitem[{Iuspa \& Scaramuzzino(2001)}]{iuspa:2001}
Iuspa, L., \& Scaramuzzino, F. 2001, Soft Computing - A Fusion of Foundations,
  Methodologies and Applications, 5, 58, 10.1007/s005000000066

\bibitem[{{Karr} \& {Freeman}(1999)}]{karr:1999}
{Karr}, C.~L., \& {Freeman}, L.~M. 1999, Industrial Applications of Genetic
  Algorithms, CRC Press International Series On Computational Intelligence (CRC
  Press)

\bibitem[{{Lagg} {et~al.}(2004){Lagg}, {Woch}, {Krupp}, \&
  {Solanki}}]{lagg:2004}
{Lagg}, A., {Woch}, J., {Krupp}, N., \& {Solanki}, S.~K. 2004, \aap, 414, 1109

\bibitem[{{Landi Degl'Innocenti}(1994)}]{landideglinnocenti:1994}
{Landi Degl'Innocenti}, E. 1994, in Solar Surface Magnetism, ed. {R.~J.~Rutten
  \& C.~J.~Schrijver}, 29

\bibitem[{{Landi Degl'Innocenti} \& {Landolfi}(2004)}]{landideglinnocenti:2004}
{Landi Degl'Innocenti}, E., \& {Landolfi}, M., eds. 2004, Astrophysics and
  Space Science Library, Vol. 307, {Polarization in Spectral Lines}

\bibitem[{{Levenberg}(1944)}]{levenberg:1944}
{Levenberg}, K. 1944, Quarterly of Applied Mathematics, 2, 164

\bibitem[{{Mahfoud} \& {Mani}(1996)}]{mahfoud:1996}
{Mahfoud}, S., \& {Mani}, G. 1996, Applied Artificial Intelligence, 10, 543

\bibitem[{{Marquardt}(1963)}]{marquardt:1963}
{Marquardt}, D.~W. 1963, SIAM Journal on Applied Mathematics, 11, 431

\bibitem[{{McIntosh} {et~al.}(1998){McIntosh}, {Diver}, {Judge}, {Charbonneau},
  {Ireland}, \& {Brown}}]{mcintosh:1998}
{McIntosh}, S.~W., {Diver}, D.~A., {Judge}, P.~G., {et~al.} 1998, \aaps, 132,
  145

\bibitem[{{NVIDIA Corp.}(2010)}]{nvidia:2010}
{NVIDIA Corp.} 2010, {Tesla C1060 Computing Processor Board: Board
  Specifications} (NVIDIA Corp.)

\bibitem[{{NVIDIA Corp.}(2011)}]{nvidia:2011}
---. 2011, {NVIDIA CUDA C Programming Guide} (NVIDIA Corp.)

\bibitem[{{Orozco Su{\'a}rez} \& {Del Toro Iniesta}(2007)}]{orozcosuarez:2007}
{Orozco Su{\'a}rez}, D., \& {Del Toro Iniesta}, J.~C. 2007, \aap, 462, 1137

\bibitem[{{Pearson}(1901)}]{pearson:1901}
{Pearson}, K. 1901, Philosophical Magazine, 2(6), 559

\bibitem[{{Pierce} \& {Breckinridge}(1973)}]{pierce:1973}
{Pierce}, A.~K., \& {Breckinridge}, J.~B. 1973, The Kitt Peak Table of
  Photographic Solar Spectrum Wavelengths, Contribution (Kitt Peak National
  Observatory) (Kitt Peak National Observatory)

\bibitem[{{Press} {et~al.}(1988){Press}, {Teukolsky}, {Flannery}, \&
  {Vetterling}}]{press:1988}
{Press}, W.~H., {Teukolsky}, S.~A., {Flannery}, B.~P., \& {Vetterling}, W.~T.
  1988, Numerical Recipes in C (Cambridge University Press)

\bibitem[{{Rachkovsky}(1962)}]{rachkovsky:1962}
{Rachkovsky}, D.~N. 1962, Izv. Krymsk. Astrofiz. Obs., 27, 148

\bibitem[{{Ramirez} \& {Fuentes}(2002)}]{federicoramirez:2002}
{Ramirez}, J., \& {Fuentes}, O. 2002, Experimental Astronomy, 14, 129,
  10.1023/B:EXPA.0000009933.44289.e4

\bibitem[{{Rees} {et~al.}(2000){Rees}, {L{\'o}pez Ariste}, {Thatcher}, \&
  {Semel}}]{rees:2000}
{Rees}, D.~E., {L{\'o}pez Ariste}, A., {Thatcher}, J., \& {Semel}, M. 2000,
  \aap, 355, 759

\bibitem[{{Rees} \& {Semel}(1979)}]{rees:1979}
{Rees}, D.~E., \& {Semel}, M.~D. 1979, \aap, 74, 1

\bibitem[{{Ronan} {et~al.}(1987){Ronan}, {Mickey}, \& {Orrall}}]{ronan:1987}
{Ronan}, R.~S., {Mickey}, D.~L., \& {Orrall}, F.~Q. 1987, \solphys, 113, 353

\bibitem[{{Ruiz Cobo} \& {del Toro Iniesta}(1992)}]{ruizcobo:1992}
{Ruiz Cobo}, B., \& {del Toro Iniesta}, J.~C. 1992, \apj, 398, 375

\bibitem[{{Sanchez Almeida}(1997)}]{sanchezalmeida:1997}
{Sanchez Almeida}, J. 1997, \apj, 491, 993

\bibitem[{{Skumanich} \& {Lites}(1987)}]{skumanich:1987}
{Skumanich}, A., \& {Lites}, B.~W. 1987, \apj, 322, 473

\bibitem[{{Socas-Navarro} {et~al.}(2001){Socas-Navarro}, {L{\'o}pez Ariste}, \&
  {Lites}}]{socasnavarro:2001}
{Socas-Navarro}, H., {L{\'o}pez Ariste}, A., \& {Lites}, B.~W. 2001, \apj, 553,
  949

\bibitem[{{Unno}(1956)}]{unno:1956}
{Unno}, W. 1956, \pasj, 8, 108

\end{thebibliography}

\end{document}